\documentclass[aps, apl, superscriptaddress,reprint]{revtex4-1}

\usepackage{graphicx}
\usepackage{bm}
\usepackage{color}
\usepackage{amsmath}
\usepackage{epstopdf}
\usepackage{amssymb}

\begin{document}

\title{Tailoring N\'eel orders in Layered Topological Antiferromagnets}

\author{Xiaotian Yang}
\thanks{These authors contribute equally to this work.}
\affiliation{School of Physical Science and Technology, ShanghaiTech University, Shanghai, 201210, China.}
\affiliation{ShanghaiTech Laboratory for Topological Physics, ShanghaiTech University, Shanghai, 201210, China.}
\author{Yongqian Wang}
\thanks{These authors contribute equally to this work.}
\affiliation{School of Physics, Renmin University of China, Beijing, 100872, China.}
\affiliation{Key Laboratory of Quantum State Construction and Manipulation (Ministry of Education), Renmin University of China, Beijing, 100872, China.}
\author{Yongchao Wang}
\affiliation{State Key Laboratory of Low Dimensional Quantum Physics, Tsinghua University, Beijing, 100084, China.}
\author{Zichen Lian}
\affiliation{State Key Laboratory of Low Dimensional Quantum Physics, Tsinghua University, Beijing, 100084, China.}
\author{Jinsong Zhang}
\affiliation{State Key Laboratory of Low Dimensional Quantum Physics, Tsinghua University, Beijing, 100084, China.}
\affiliation{Frontier Science Center for Quantum Information, Beijing, 100084, China.}
\affiliation{Hefei National Laboratory, Hefei, 230088, China.}
\author{Yayu Wang}
\affiliation{State Key Laboratory of Low Dimensional Quantum Physics, Tsinghua University, Beijing, 100084, China.}
\affiliation{Frontier Science Center for Quantum Information, Beijing, 100084, China.}
\affiliation{Hefei National Laboratory, Hefei, 230088, China.}
\affiliation{New Cornerstone Science Laboratory, Frontier Science Center for Quantum Information, Beijing, 100084, China.}
\author{Chang Liu}
\email[Corresponding authors:\\]{liuchang\_phy@ruc.edu.cn, wangwb1@shanghaitech.edu.cn}
\affiliation{School of Physics, Renmin University of China, Beijing, 100872, China.}
\affiliation{Key Laboratory of Quantum State Construction and Manipulation (Ministry of Education), Renmin University of China, Beijing, 100872, China.}
\author{Wenbo Wang}
\email[Corresponding authors:\\]{liuchang\_phy@ruc.edu.cn, wangwb1@shanghaitech.edu.cn}
\affiliation{School of Physical Science and Technology, ShanghaiTech University, Shanghai, 201210, China.}
\affiliation{ShanghaiTech Laboratory for Topological Physics, ShanghaiTech University, Shanghai, 201210, China.}

\maketitle

\textbf{In the two-dimensional limit, the interplay between N\'eel order and band topology in van der Waals topological antiferromagnets can give rise to novel quantum phenomena in the quantum anomalous Hall state, including the cascaded quantum phase transition and spin-modulation effect\cite{Deng2020,Fu2020,Li2021,Bai2023,Lian2024}. However, due to the absence of net magnetization in antiferromagnets, probing the energetically degenerate N\'eel orders has long remained a significant challenge. Inspired by recent advances in realizing the quantum anomalous Hall effect in AlO$_x$-capped layered topological antiferromagnet MnBi$_2$Te$_4$\cite{Lian2024,Li2024,Wang2025}, we demonstrate deterministic control over the N\'eel order through surface anisotropy engineering enabled by the AlO$_x$ capping layer. By tuning the surface anisotropy, we uncover parity-dependent symmetry breaking states that manifest as distinct odd-even boundary architectures, including 180$^\circ$ domain walls or continuous spin structures. Comparative studies between AlO$_x$-capped and pristine odd-layer MnBi$_2$Te$_4$ flakes using domain-resolved magnetic force microscopy reveal pronounced differences in coercivity and magnetization-reversal dynamics. Notably, an unconventional giant exchange bias, which arises from perpendicular magnetic anisotropy rather than traditional interface pinning mechanisms, is observed for the first time. Our findings establish a pathway for manipulating N\'eel order through surface modification in A-type antiferromagnets, offering new opportunities for spintronic devices and quantum information technologies.}

\begin{figure*}[htp]
\includegraphics[width=0.85\textwidth]{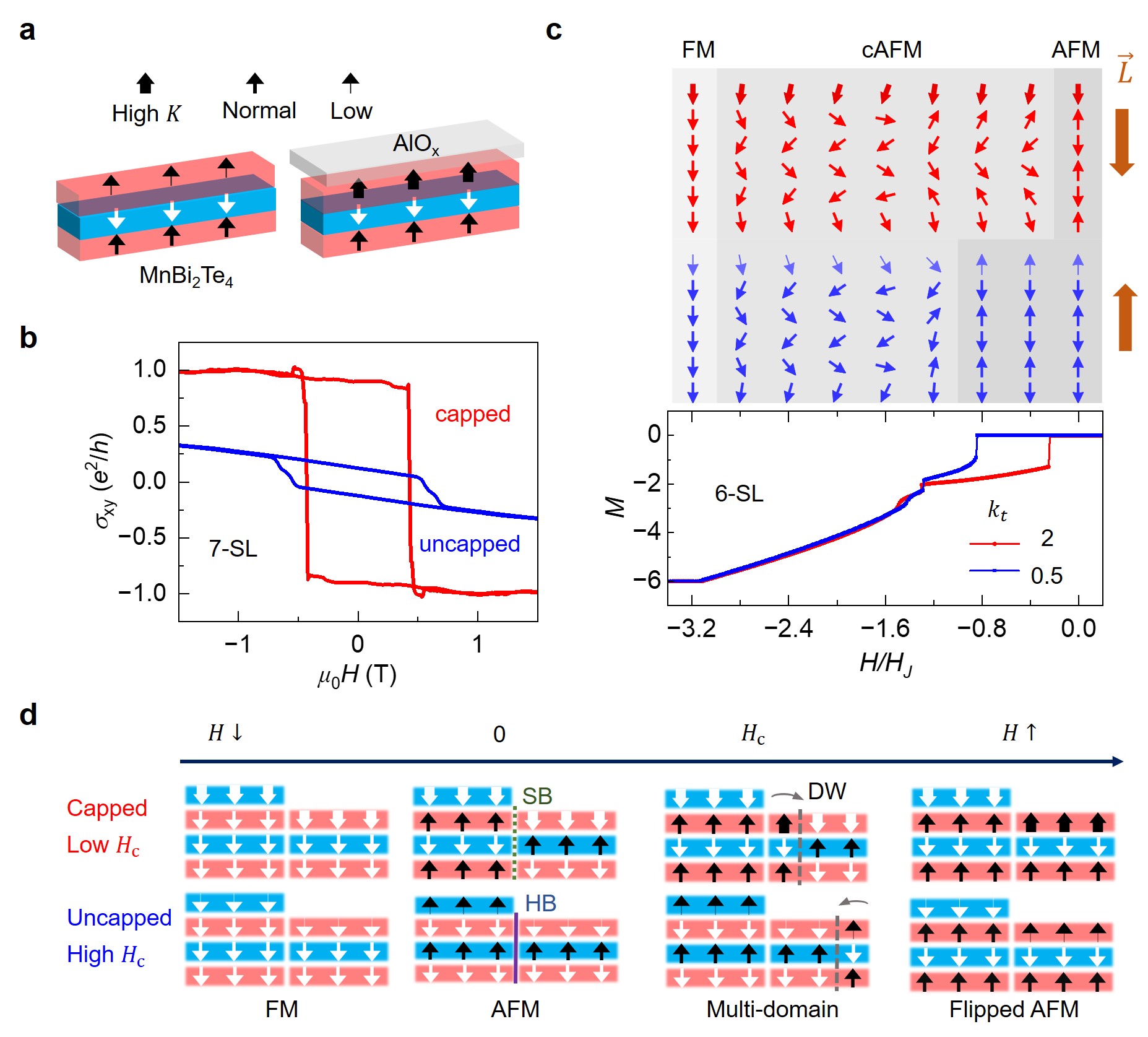}
\caption{\textbf{N\'eel orders of AlO$_x$-capped and pristine MBT thin flakes} $\textbf{a}$, Schematic representation of enhanced (thicker arrows) and suppressed (thinner arrows) surface magnetic anisotropy in  AlO$_x$-capped and pristine MnBi$_2$Te$_4$. \textbf{b}, Field-dependent Hall conductance ($\sigma_{xy}$) of capped (1.5\,K) and uncapped (2\,K) 7-SL MnBi$_2$Te$_4$ at charge neutral points. \textbf{c}, Simulated spin reorientations for 6-SL systems upon field removal, showing metamagnetic transition from ferromagnetic (FM) to canted-antiferromagnetic (cAFM) to antiferromagnetic (AFM) phases. Red spins are for enhanced surface anisotropy ($k_t=2$), while blue ones are for suppressed surface anisotropy ($k_t=0.5$). The N\'eel vectors $\boldsymbol{L}$ of the AFM ground states at zero fields show opposite directions. Corresponding $M$-$H$ curves are plotted in lower panel. \textbf{d}, Schematic illustrations of the magnetization reversal process of odd-even staircases for enhanced (capped) and suppressed (uncapped) surface anisotropy. The magnetically soft and hard OEBs are labelled as SB and HB, respectively, resulting in lower and higher coercive field ($H_c$). The domain wall (DW) propagation directions differ between these two cases.
\label{fig1}}
\end{figure*}

\begin{figure*}[htp]
\includegraphics[width=0.8\textwidth]{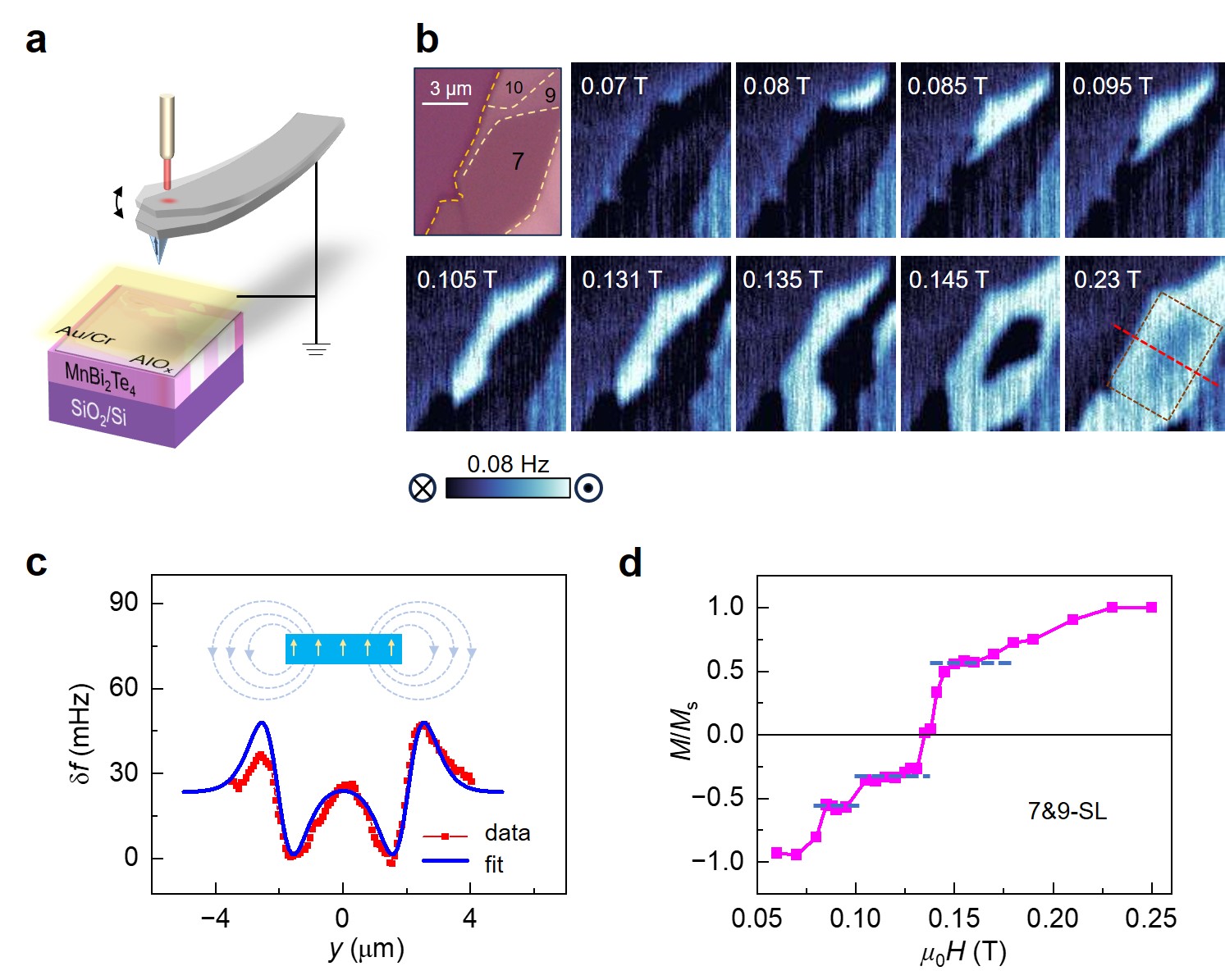}
\caption{\textbf{Robust A-type antiferromagnetism in MBT thin flakes.} $\textbf{a}$, Schematic illustration of MFM setup based on interferometer sensors, scanning on AlO$_x$-capped MnBi$_2$Te$_4$ staircases. The sample was coated with Cr and Au metals, which was utilized as electrode to balance electric potential between the magnetic tip and sample surface. \textbf{b}, MFM measurements on AlO$_x$-capped MnBi$_2$Te$_4$ thin flakes at various magnetic fields. The sample edge is marked by yellow dashed lines, while atomic steps on the thin flake are outlined by light-yellow dashed lines. The number of SLs is labelled on the corresponding terraces. \textbf{c}, Line profile (red dashed line) of the MFM images at 0.23\,T, which can be accurately fitted by the dipole field from a uniformly magnetized rectangular sheet (yellow dashed rectangle). \textbf{d} Normalized magnetization of even-SL thin flakes, deduced from MFM images at various fields, exhibits Barkhausen effect. Small magnetization plateaus are indicated by blue dashed lines.
\label{fig2}}
\end{figure*}

\begin{figure*}[htp]
\includegraphics[width=0.9\textwidth]{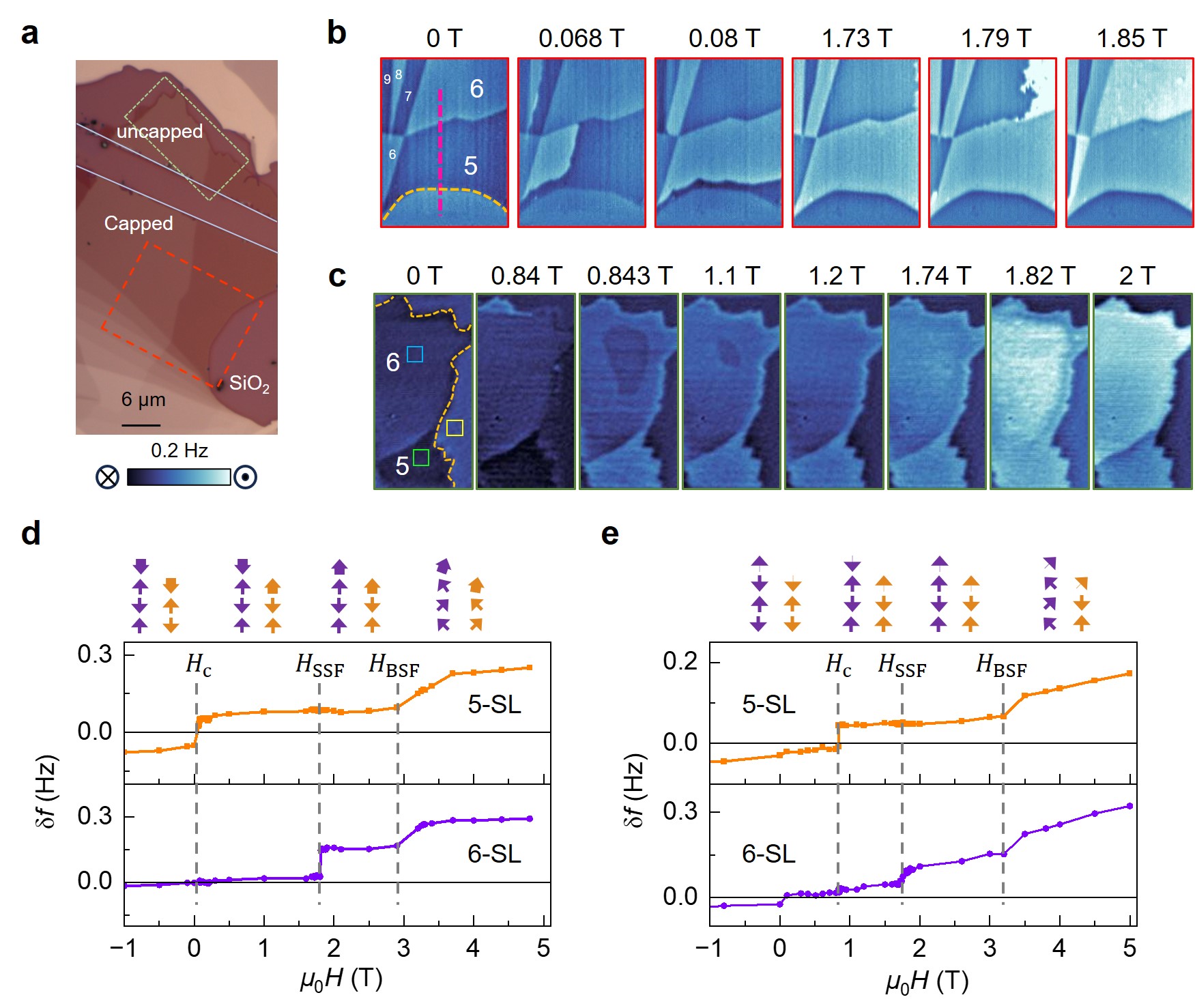}
\caption{\textbf{Distinct magnetization reversal processes of capped and pristine MBT thin flakes.} $\textbf{a}$, Optical image of the AlO$_x$-capped and uncapped MBT from the same specimen. The scratch separating capped and uncapped regions are labelled by blue solid lines, and the MFM scanning areas are marked by red and green dashed lines. \textbf{b}, MFM images of capped MBT at various magnetic fields. The flake edge is outlined by yellow dashed lines. \textbf{c}, MFM images of uncapped MBT at various magnetic fields. \textbf{d}, MFM signals of 5- and 6-SL capped MBT, derived from the domain contrast along the line profile shown in \textbf{b}. \textbf{e},  MFM signals of 5- and 6-SL uncapped MBT, derived from the domain contrast within the square areas shown in  \textbf{c}. Critical magnetic fields for different magnetic phases are indicated by gray dashed lines. $H_\mathrm{c}$, $H_\mathrm{SSF}$ and $H_\mathrm{BSF}$ represent the coercive field of 5-SL, the surface-spin-flip field of 6-SL, the bulk-spin-flop field of both thin flakes. The schematic spin structures corresponding to each magnetic phase are illustrated at the top of the graphs.
\label{fig3}}
\end{figure*}

A-type antiferromagnetic materials exhibit a dual magnetic character defined by intralayer ferromagnetic coupling and interlayer antiferromagnetic interactions\cite{May2009,Baltz2018,Gong2018,Jiang2018,Otrokov2019,Yan2020,Rahman2021,Ye2022,Schrunk2022,chen2024}. This unique magnetic order manifests distinct layer-parity-dependent phenomena in few-layer thin flakes: even-layer systems with zero net moment behave as conventional antiferromagnets, whereas odd-layer configurations retain a net magnetic moment akin to ferromagnets. As the first experimentally identified layered antiferromagnet with nontrivial band topology, MnBi$_2$Te$_4$ (MBT) family materials offer a versatile platform to explore the intricate interaction among dimensionality, topology and antiferromagnetism. In MBT, the parity-dependent magnetism is directly coupled to its topological electronic states. Even-layer systems stabilize axion insulator phases\cite{Zhang2019,Li2019,Liu2020}, while odd-layer counterparts host quantum anomalous Hall (QAH) states\cite{Deng2020,Fu2020,Li2021,Bai2023}. Precise manipulation of magnetic spin states has become essential for topological systems, given their direct correlation with topological invariants\cite{Mong2010,Li2019a,Chen2019,Hu2020,Ma2020,Gao2021,Huan2021}. However, the magnetic ground state of A-type antiferromagnets comprises a two-fold degeneracy between antiparallel N\'eel orders, which is difficult to detect in conventional magnetic measurements. For odd-layer systems, the N\'eel vector aligns with the net magnetization, enabling strong coupling to external magnetic fields. In contrast, even-layer systems without net magnetization exhibit field-insensitive N\'eel orders, making it challenging to artificially control the N\'eel states. 

Recent advances in interfacial engineering, such as AlO$_x$ capping of MnBi$_2$Te$_4$ surfaces, suggest a pathway to address this challenge. AlO$_x$ contact with either surface of MBT significantly amplifies the anomalous Hall effect toward full quantization, likely attributed to the enhanced surface perpendicular magnetic anisotropy (PMA)\cite{Deng2020,Zhang2022,Li2024,Wang2025,Lian2024}. This interfacial engineering strategy, which involves coating magnetic systems with amorphous oxide capping layers to stabilize PMA, has been experimentally implemented in diverse systems\cite{Monso2002,Rodmacq2003,Dieny2017}. In pristine MBT systems, surface magnetic parameters--including magnetic moment, exchange coupling, and magnetic anisotropy--are typically weaker than those in the bulk\cite{Bing2021}. This disparity manifests as multiple-step metamagnetic transitions during magnetization reversal \cite{Sass2020,Ge2022}. Capping a layer of AlO$_x$ can enhance surface PMA, potentially surpassing even the bulk PMA (Fig.~\ref{fig1}\textbf{a}). This effect effectively pins the surface spin direction, enabling deterministic control over the N\'eel orders in even-layer systems.

In this study, we introduce a methodology to determine and manipulate N\'eel orders in A-type antiferromagnets through surface modification. In AlO$_x$-capped and pristine MnBi$_2$Te$_4$ thin flakes, the odd-even boundaries (OEB) exhibit distinct magnetic properties during magnetization reversal, indicating inverted N\'eel orders in the even-layer regions. The unique coercivity of OEBs can be leveraged to design giant exchange bias effects in odd-layer thin flakes. This design strategy extends beyond MnBi$_2$Te$_4$ systems and is applicable to other A-type van der Waals antiferromagnets, offering a versatile approach for controlling antiferromagnetic states in layered materials.

\subsection*{Transport and calculated N\'eel orders}\label{sec2}

To validate the role of AlO$_x$ capping, we fabricated both AlO$_x$-capped and pristine MnBi$_2$Te$_4$ thin flakes and systematically characterized their Hall conductances. At 1.5\,K, the $\sigma_{xy}$-$H$ loops for AlO$_x$-capped 7-septuple-layer (7-SL) devices exhibit square-like hysteretic behavior with a zero-field Hall conductance reaching approximately $0.90e^2/h$, approaching the quantum anomalous Hall (QAH) state, as illustrated in  Fig.~\ref{fig1}\textbf{b}. The temperature dependence of the anomalous Hall conductance aligns well with the mean-field theory for a Landau second-order phase transition, described by $\sigma_{xy} \propto(T_N-T)^{0.5}$, where $T_N \approx 22.0$\,K, highlighting the robustness of A-type antiferromagnetic ordering\cite{Chang2015,Ou2018}. In contrast, the uncapped flake shows a significantly lower anomalous Hall conductance, consistent with prior findings that suggest AlO$_x$ capping enhances PMA\cite{Li2024,Wang2025}. 

A linear-chain model was employed to calculate how surface magnetic anisotropy modifies the spin states at various magnetic fields (methods)\cite{Mills1968,Yang2021,Ovchinnikov2021,Chong2024}. Each layer was treated as a homogeneous macro moment $\boldsymbol{m}$. The hysteretic metastable states were determined by minimizing the total magnetic free energy, which includes contributions from  Zeeman energy, PMA energy, and interlayer antiferromagnetic exchange energy. At high negative fields, all moments align downward, forming a ferromagnetic (FM) state. As the field is reduced to zero, the system transitions through a canted antiferromagnetic (cAFM) state and finally reaches the antiferromagnetic (AFM) ground state. The N\'eel vectors of the AFM ground states are defined as $\boldsymbol{L}=\boldsymbol{m}_1-\boldsymbol{m}_2$, where $\boldsymbol{m}_1$ and $\boldsymbol{m}_2$ represent the macro moment of the surface and subsurface layer, respectively. During the metamagnetic transition of even-SL systems, enhanced surface PMA pins the spin orientation of the topmost layer, thereby stabilizing an AFM ground state with a downward N\'eel vector at zero field. Conversely, suppressed surface PMA triggers spin reorientation in the surface layer, ultimately stabilizing an upward N\'eel vector. As shown in Fig.~\ref{fig1}\textbf{c}, a doubled surface anisotropy in 6-SL flake results in a $\downarrow\uparrow\downarrow\uparrow\downarrow\uparrow$ moment sequence (red), while halved surface anisotropy yields $\uparrow\downarrow\uparrow\downarrow\uparrow\downarrow$ (blue). In contrast, odd-SL flakes, regardless of the surface anisotropy, exhibits the identical AFM states, such as $\downarrow\uparrow\downarrow\uparrow\downarrow$, ensuring a net downward magnetization during metamagnetic transitions. Consequently, OEBs exhibit distinct magnetic properties at zero fields. Enhanced surface anisotropy results in a 180$^\circ$ domain wall, while reduced surface anisotropy maintains coherent spin continuity. When a positive field is applied, the N\'eel vector of odd-SL flakes flips, transitioning from $\downarrow\uparrow\downarrow$ to $\uparrow\downarrow\uparrow$. The magnetization reversal processes at these two types of OEBs differ significantly (Fig.~\ref{fig1}\textbf{d}). For a 180$^\circ$ domain wall, reversal occurs through domain-wall propagation without domain nucleation. In capped MBT, magnetization switching initiates at OEBs, and the coercivity is negligible if domain-wall pinning is weak. This OEB is termed a magnetically soft boundary (SB), which is marked by green vertical dashed line in Fig.~\ref{fig1}\textbf{d}. Conversely, uncapped flakes tend to form a single-domain state with continuous spin structures across the OEB, representing a magnetically hard boundary (HB, marked by purple vertical solid line), which makes magnetization switching energetically unfavorable. Domain switching in this case requires overcoming a domain nucleation barrier, which scales with magnetic anisotropy energy $K$\cite{Brown1945,Aharoni1962}. Consequently, domain switching in uncapped flakes initiates at sample edges or involves coherent spin flipping in the absence of domain formation. Theoretically, the coercivity difference between these two scenarios is $\mu_0\Delta H_\mathrm{c}\approx K/M_s$, where $M_s$ is the saturation magnetization.

\subsection*{Domain behavior during spin flips}\label{sec3}

In order to experimentally verify this prediction, we utilized cryogenic magnetic force microscopy (MFM) to probe the magnetic domain evolutions at various magnetic fields (Fig.~\ref{fig2}\textbf{a}). Magnetic imaging was first performed on AlO$_x$-capped MBT thin flakes, focusing on the 7-SL terrace, where QAHE has been realized. The thin flake had been magnetically saturated at negative fields, then the field was removed and ramped to positive. The magnetization reversal process of the odd-layer terraces exhibits a typical ferromagnetic domain behavior with domain wall propagation, as shown in Fig.~\ref{fig2}\textbf{b}. At 0.08\,T, up domains started to nucleate from the 9-/10-SL boundary and domain wall propagated into the 7-SL region. The magnetization switching gradually occurred at the edges of the 7-SL terrace, with the center area unswitched at 0.145\,T. Domain patterns exhibit remarkable stability across the field intervals of 0.085$\sim$0.095\,T and 0.105$\sim$0.131\,T, due to domain wall pinning effects. The odd-SL area was completed switched at 0.23\,T. A line cut across the 7-SL terrace reveals the local dependence of the MFM signals, given by $\delta f \propto \frac{d^2B_z}{dz^2}$ for a magnetic dipole tip\cite{Hug1998}. The magnetic field $B_z$ generated by a ferromagnetic thin flake with PMA is analogous to the electric field outside a parallel capacitor. The dipole field can be derived from the scalar potential $\phi$: $B_z\propto-\frac{d\phi}{dz}$. For a rectangular thin flake the scalar potential across the midline of the long edge can be expressed as $\phi(z) \propto \arctan \frac{L(W-2y)}{2z\sqrt{L^2+(W-2y)^2+4z^2}}+\arctan \frac{L(W+2y)}{2z\sqrt{L^2+(W+2y)^2+4z^2}}$,
where $L=5.1$\,$\mu$m, $W=4.1$\,$\mu$m are respectively the length and width of the rectangle\cite{Parker2002}.
As shown in Fig.~\ref{fig2}\textbf{c}, the line profile can be nicely fitted by this model at $z=1.2$\,$\mu$m, suggesting the effective lift height of the tip dipole moment. This consistency indicates the homogeneous magnetization in the MBT thin flake. By calculating the switched areas, the normalized magnetization $M/M_\mathrm{s}$ can be estimated at various fields. Fine magnetization plateaus and Barkhausen jumps are evident on the magnetization switching curve in Fig.~\ref{fig2}\textbf{d}, highlighting robust intralayer ferromagnetic ordering\cite{Kim2003,Schwarz2004}. This magnetization reversal process starkly contrasts with other QAH systems, including magnetically doped topological insulator films\cite{Lachman2017,Wang2018}, MBT films\cite{Shi2024} and twisted bilayer systems\cite{Tschirhart2021,Redekop2024}, where disordered magnetic domains or even superparamagnetic behaviors have been visualized, indicating magnetic inhomogeneity.

\newpage
\begin{figure*}[htp]
\includegraphics[width=1\textwidth]{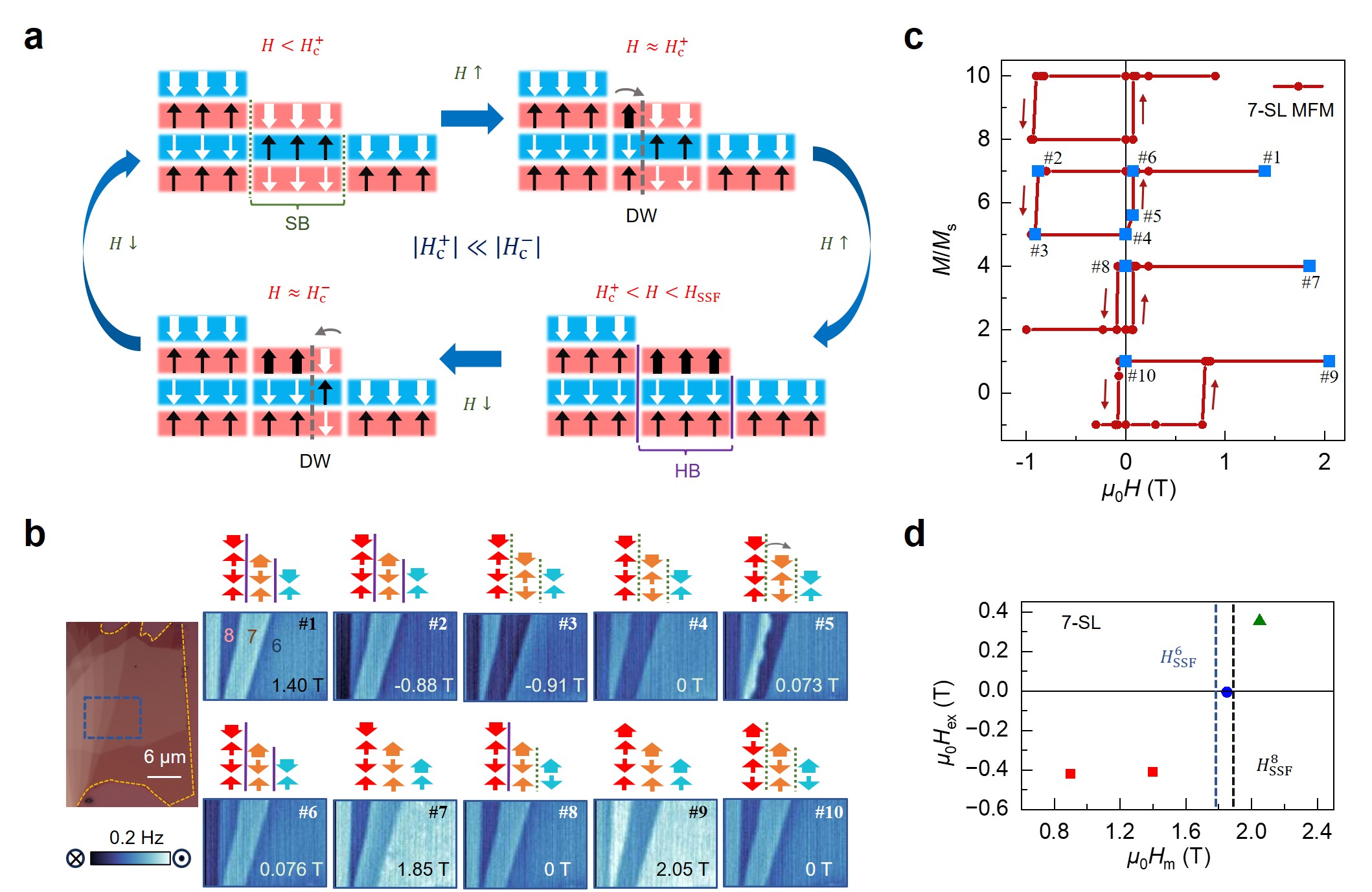}
\caption{\textbf{Giant exchange bias in odd-SL MBT encircled by even-SL terraces} $\textbf{a}$, Schematic diagrams of exchange bias effect in 2-/3-/4-SL staircase structures, when the applied field is below the surface-spin-flip field ($H_\mathrm{SSF}$) of the even-SL. The SBs, HBs, DWs are labelled by green dashed, purple solid and gray dashed lines, respectively. The positive coercive field ($H_\mathrm{c}^+$) is much smaller than negative one ($H_\mathrm{c}^-$). $\textbf{b}$, Optical image of the thin flake with its edges outlined by yellow dashed lines. MFM scanning area is highlighted by blue dashed lines. MFM images of 6-/7-/8-SL staircase structure at various magnetic fields (\#1-10). The corresponding spin structures are illustrated at top. $\textbf{c}$, Magnetization curves of 7-SL MBT derived from MFM images exhibit exchange bias effect. Red circles are data points from MFM images. Blue squares correspond to the representative MFM images shown in $\textbf{b}$. The red arrows represent the field-sweeping directions. $\textbf{d}$, Exchange fields $H_\mathrm{ex}$ for distinct maximum positive fields ($H_\mathrm{m}$) demonstrate strong correlation with the spin states of neighboring even-SL regions. $H^6_\mathrm{SSF}$ and $H^8_\mathrm{SSF}$ denote surface-spin-flip fields for 6-SL and 8-SL terraces, respectively.
\label{fig4}}
\end{figure*}

To directly compare the impact of AlO$_x$ capping on surface magnetism, we cleaved another MBT thin flake containing regions of varying thicknesses. Half of the flake was capped with a 3\,nm AlO$_x$ layer, with the AlO$_x$ boundary intentionally aligned across the interface between 5- and 6-SL regions (Fig.~\ref{fig3}\textbf{a}). The flake was then mechanically split along the AlO$_x$ boundary using a sharp needle. Subsequent MFM imaging was performed on both AlO$_x$-capped and uncapped 5- and 6-SL regions. In capped regions, magnetization reversal initiated in the 5-SL segment at 0.068\,T, with domain nucleation observed precisely at the 5-/6-SL boundary (Fig.~\ref{fig3}\textbf{b}), analogous to previous reversal dynamics in 7-SL systems. Upon increasing the field to 1.73\,T, the 6-SL region underwent magnetization reversal, attributed to surface spin flip (SSF), accompanied by domain formation. The propagation from right to left of the magnetic domain walls stems from the presence of an AlO$_x$ capping boundary at the right edge of the sample. This directional motion provides direct evidence that 6-SL SSF transitions nucleate preferentially within uncapped regions, where reduced surface PMA lowers the energy barrier for spin flips. At 1.85\,T, the SSF transition in the 6-layer region completed, and further field increases triggered bulk magnetization reversal in both 6- and 5-SL regions. Notably, this bulk spin flip (BSF) occurred without observable domain nucleation, suggesting a coherent spin-flop process across the entire lattice. By analyzing the MFM signal contrast between the 5-/6-SL and the 5-SL/substrate boundaries, we reconstructed layer-resolved magnetic hysteresis curves. These curves reveal distinct critical fields: $\mu_0 H_\mathrm{c}\approx 0.068$\,T for 5-SL spin flip, $\mu_0 H_\mathrm{SSF} \approx 1.80$\,T for 6-SL SSF, and $\mu_0 H_\mathrm{BSF} \approx 2.90$\,T for BSF in both SLs (Fig.~\ref{fig3}\textbf{d}). In contrast, uncapped flakes exhibit no magnetization reversal below 0.84\,T. At 0.843\,T, abrupt spin-flip transitions occur: complete reversal in the 5-SL and partial reversal in the 6-SL, leaving a metastable unswitched core (Fig.~\ref{fig3}\textbf{c}). This behavior starkly contrasts with AlO$_x$-capped systems. The absence of domain nucleation at the 5-/6-SL boundary and the monodomain-like reversal pathway in entire uncapped flakes imply spontaneous single-domain formation at zero field, consistent with theoretical predictions of reduced surface PMA. Such layer-sharing spin-flip transition was also visualized in few-layer CrSBr systems with in-plane magnetic anisotropy\cite{Sun2025}. The disparity between capped and uncapped systems $\mu_0 \Delta H_\mathrm{c} \approx 0.77$\,T enables quantitative estimation of nucleation energy barrier. Nucleation energy barrier per Mn ion equals $ \mu_0 H_\mathrm{c}{\mu_\mathrm{Mn}}\approx0.205$\,meV, given Mn ion moment $\mu_\mathrm{Mn} \approx4.61\mu_{\mathrm{B}}$. This value is comparable with the theoretical magnetic anisotropy energy 0.225\,meV from first-principle calculations\cite{Otrokov2019}. 
This layer-parity-dependent phenomenon was further validated in a separate sample at the 7-/8-SL boundary. In AlO$_x$-capped regions, magnetization reversal nucleates preferentially at the 7-/8-SL boundary, whereas in uncapped regions, reversal initiates at the 7-SL flake edge. Furthermore, these two configurations exhibit an order-of-magnitude disparity in coercive fields. These findings conclusively demonstrate that AlO$_x$ capping amplifies surface PMA, stabilizing zero-field domain walls at OEBs.

\subsection*{Giant exchange bias effect}\label{sec4}

The stark layer-parity dependence of coercive fields enables the design of giant EB effect through tailored multilayer antiferromagnetic architectures, as illustrated in Fig.~\ref{fig4}\textbf{a}. Consider a central odd-layer terrace (e.g., 3-SL) flanked by even-layer regions (4- and 2-SL), forming a stepped structure. During field cycling from high negative fields to zero, spin configurations evolve as follows: The 4-3 and 3-2 boundaries initially act as SBs, permitting low-energy magnetization switching. The magnetization reversal prefers to initiate at the 4-3 OEBs, because the domain wall depinning at 3-2 OEBs requires the extension of domain walls which costs additional energy. Upon magnetization reversal of the 3-SL, from $\downarrow\uparrow\downarrow$ to $\uparrow\downarrow\uparrow$, these OEBs transform into HBs, requiring substantially higher negative fields to reverse the 3-SL magnetization---a hallmark of layer-correlated exchange bias. To validate this mechanism, we probed magnetization reversal in AlO$_x$-capped 6-/7-/8-SL staircase. The thin flakes were initially polarized at $-$5\,T. The field consequently were swept back and forth to obtain the $M$-$H$ loops for 7-SL. The maximum positive fields $H_m$ were increasing for each loop. For $H_m=0.90$ and 1.40\,T, lower than the even-layer $H_\mathrm{SSF}$, a constant exchange bias field was yielded in 7-SL $H_\mathrm{EB}\approx-0.41$\,T. The magnetization reversal processes for  $H_m=1.40$\,T were shown in Fig.~\ref{fig4}\textbf{c}-\textbf{h}. The field was firstly cycled to 1.40\,T, reversing the field leads to magnetization switching around $H_\mathrm{c}^-\approx-0.90$\,T (Fig.~\ref{fig4}\textbf{d},\textbf{e}). The magnetization reversal at positive field was order-of-magnitude smaller (Fig.~\ref{fig4}\textbf{g},\textbf{h}). Domain switching was visualized at $H_\mathrm{c}^+\approx0.073$\,T from 7-/8-SL OEBs. At 1.85\,T, the 6-SL surface flips while the 8-SL remains unchanged. Removing the field creates asymmetric boundaries: a soft 6-/7-SL boundary and a hard 7-/8-SL boundary (Fig.~\ref{fig4}\textbf{j}). This configuration reduces both positive and negative coercive fields ($H_\mathrm{c}^-=-0.085$\,T, $H_\mathrm{c}^+=0.074$\,T), as either magnetization state benefits from one soft nucleation boundary, effectively nullifying $H_\mathrm{EB}$. Finally, at 2.05\,T, 8-SL SSF resets the system to its initial symmetric OEB configuration (Fig.~\ref{fig4}\textbf{l}) but with inverted EB polarity $H_\mathrm{EB}\approx0.36$\,T, as shown in Fig.~\ref{fig4}\textbf{m}. Analogous EB effect with comparable $H_\mathrm{EB}$ were experimentally resolved in adjacent 9-SL regions, attributing their origin to PMA rather than thickness-dependent bulk interactions. The tight correlation between odd-SL $H_\mathrm{EB}$ and even-SL spin states unambiguously attributes this EB effect to OEBs. This observed EB mechanism is fundamentally distinct from previously reported effects in MBT. Earlier reports attributed exchange bias to defect-mediated spin pinning and inhomogeneous magnetic interactions within odd-SL regions\cite{Yang2024,Chong2024,Chen2024a}. In MBT, where the topological Chern number sign is governed by magnetization orientation, this layer-parity-mediated exchange bias offers a deterministic pathway to tailor topological edge states. 

In summary, we introduce a methodology to resolve and manipulate N\'eel orders in A-type antiferromagnets by analyzing domain-reversal processes at parity-engineered layer boundaries. In systems where surface PMA is enhanced or suppressed, parity boundaries exhibit orders-of-magnitude differences in coercive fields, serving as a magnetic fingerprint for surface spin reconfiguration. In AlO$_x$-capped MBT, this principle enables giant exchange bias effects ($\sim$0.4\,T) in odd-layer terraces. Critically, this layer-parity-engineering strategy can be extended to other A-type van der Waals antiferromagnets by artificially constructing OEBs. Unlike conventional exchange bias mechanisms, which rely on interfacial pinning in ferromagnet/antiferromagnet heterostructures and are influenced by extrinsic factors such as interface roughness and grain size, our approach leverages intrinsic parity boundaries in A-type antiferromagnetic single crystals. These naturally occurring OEBs allow for precisely tunable bias fields in odd-layer systems by manipulating the magnetic states of even-layer neighbors. Furthermore, exchange bias magnitudes, which scale with PMA, can be tailored through surface anisotropy modulation, offering a pathway to design nonvolatile spin-texture memories and topological quantum devices.

\newpage

\section*{Methods}\label{sec5}

\subsection*{Device fabrication}
Few-layer MnBi$_2$Te$_4$ flakes were mechanically exfoliated onto 285-nm-thick SiO$_2$/Si substrates using the Scotch tape method in an Ar-filled glovebox, maintaining O$_2$ and H$_2$O levels below 0.1\,ppm. To ensure a clean surface and enhance adhesion, the substrates were sequentially cleaned with acetone, isopropanol, and deionized water, followed by air plasma treatment at 125\,Pa for 3\,min. To achieve selective AlO$_x$ coverage, a physical mask was placed over the MnBi$_2$Te$_4$ flakes before aluminum deposition to partially shield specific regions. Aluminum was thermally evaporated at a deposition rate of 0.04\,nm/s under a vacuum better than $4 \times 10^{-4}$\,Pa, followed by oxidation in an oxygen atmosphere at a pressure of $2 \times 10^{-2}$\,Pa for 5\,min to form a 3-nm-thick AlO$_x$ layer in the exposed areas. After oxidation, the mask was removed, and the sample was then mechanically separated using a sharp needle to create two distinct regions with and without AlO$_x$ capping layer. To balance the electrostatic potential for subsequent MFM measurements, a Cr/Au (3/25\,nm) layer was deposited via thermal evaporation, ensuring uniform electrical contact across both regions and minimizing electrostatic artifacts that could interfere with magnetic imaging.

\subsection*{Transport measurements}

Four-probe transport measurements were performed in a cryogenic system operating at a base temperature of 1.5\,K, with an out-of-plane magnetic field up to 9\,T. An AC current of 100\,nA at 13\,Hz, generated by a Keithley 6221 current source, was applied to the sample, and the longitudinal and Hall voltage were simultaneously detected using lock-in amplifiers (Stanford Research Systems SR830). To correct for geometric misalignment, the longitudinal and Hall signals were symmetrized and anti-symmetrized with respect to the applied magnetic field. The back-gate voltage was controlled via a Keithley 2400 source meter through the SiO$_2$/Si substrate.

\subsection*{Linear-chain model calculations}
In A-type antiferromagnetic materials, we can regard $i$th layer as a macro moment $\boldsymbol{m}_i$. For homogeneous macro-moment size, we can set $|\boldsymbol{m}_i|=1$ for simplicity. The toal magnetic free energy of N-SL MnBi$_2$Te$_4$ can be written as 
\begin{equation}
\resizebox{1\hsize}{!}{$E= -H\sum_{i=1}^{N} \cos\theta_i+H_K(\sum_{i=2}^{N} \sin^{2}\theta_i+k_t\sin^{2}\theta_1)
+H_J\sum_{i=1}^{N-1}\cos(\theta_{i}-\theta_{i+1})$}
\end{equation}
where $H$ is magnetic field along $z$ axis and $\theta_i$ is the angle between $\boldsymbol{m}_{i}$ and $z$ axis. These three terms correspond to Zeeman energy, PMA energy ($H_K>0$) and interlayer antiferromagnetic exchange energy ($H_J>0$), respectively. $k_t$ term represents the coefficient of the surface PMA. For MBT, we set $H_K/H_J=0.3$\cite{Yang2021}. The spin orientations \{$\theta_i$\} therefore can be determined by searching for the nearest local minimum evolving from the previous state. In $\textit{mathematica}$, we utilized $\textit{FindMinimum}$ to efficiently find the local minimum of total free energy. The magnetization curve can be obtained by calculating $M=\sum_{i=1}^{N} \cos\theta_i$.

\subsection*{MFM measurements}

 Cryogenic MFM experiments were conducted in a commercial atomic force microscope (atto-AFM) equipped with commercial cantilevers (spring constant $k \approx$2.8\,N/m and resonance frequency $\approx$75.8\,kHz) in a closed-cycle helium cryostat. The tip deflections were detected by an all-fiber low-coherence interferometer. An out-of-plane magnetic field was applied using a superconducting magnet. MFM images were taken in a constant height mode with lift height of $\sim$200nm. MFM signal, the change of cantilever resonance frequency, is proportional to the gradient of out-of-plane stray field. Electrostatic interaction was minimized by balancing the tip-surface potential difference. To avoid the relaxation effect (domain wall creeping) during the spin-flip transitions of odd-SL MBT near $H_\mathrm{c}$ and to minimize the stray field effect of the MFM tips, all MFM images were taken at a low magnetic field of approximately 0.03\,T after the magnetic field was ramped to the desired values. 
 
\vskip 0.5cm
\noindent {\bf Acknowledgements }We thank Y. Chen and Z. Zhu for discussions. Wenbo Wang was sponsored by National Key Research and Development Program of China (Grant No. 2022YFA1403000), and National Natural Science Foundation of China (Grant No. 12374161). Yayu Wang was supported by the Basic Science Center Project of Natural Science Foundation of China (Grant No. 52388201), the Innovation program for Quantum Science and Technology (Grant No. 2021ZD0302502) and the New Cornerstone Science Foundation through the New Cornerstone Investigator Program and the XPLORER PRIZE. Chang Liu was supported by the National Natural Science Foundation of China (Grant No. 12274453); Beijing Nova Program (Grant No. 20240484574). Jinsong Zhang was supported by the National Natural Science Foundation of China (Grant Nos. 12274252 and 12350404).

\vskip 0.5cm
\noindent {\bf Author contributions }W.B.W conceived the project. W.B.W., Y.Y.W, C.L and J.S.Z  supervised the research. Y.C.W. synthesized the MnBi$_2$Te$_4$ single crystals. Y.Q.W. and Z.C.L. fabricated the devices and performed the transport measurements. X.T.Y. and W.B.W performed the MFM experiments and analyzed the data. W.B.W. performed the linear-chain model calculations. W.B.W. and X.T.Y. wrote the manuscript. All authors discussed the data and contributed to the manuscript.

\newpage

\end{document}